\documentclass[a4paper,twocolumn,preprintnumbers,superscriptaddress,citeautoscript,newabstract,aps,nofootinbib]{revtex4-2} 
\usepackage{amsmath,amsfonts,amssymb,graphics}
\usepackage[normalem]{ulem}
\usepackage{graphicx}
\usepackage{units}
\usepackage{mathrsfs}
\usepackage{bbm}
\usepackage{pifont}
\usepackage{braket}
\usepackage{units}
\usepackage[dvipsnames]{xcolor}
\usepackage{mathtools}
\usepackage{xspace}
\usepackage{booktabs}

\usepackage[bookmarks=false,hyperfootnotes=false]{hyperref}

\definecolor{nicered}{rgb}{0.5,.0,.0}
\definecolor{darkblue}{rgb}{0,.1,.9}
\definecolor{lightblue}{rgb}{0,.1,.6}
\definecolor{darkgreen}{rgb}{0.0,0.2,0.0}
\hypersetup{colorlinks, citecolor=darkgreen ,linkcolor=darkblue, urlcolor=lightblue}

\newcommand{\SO}[1]{\ensuremath{\mathrm{SO}(#1)}}
\newcommand{\SU}[1]{\ensuremath{\mathrm{SU}(#1)}}

\newcommand{\U}[1]{\ensuremath{\mathrm{U}(#1)}}

\newcommand{\rep}[2][]{\ensuremath{\boldsymbol{#2}#1}}

\renewcommand{\bar}[1]{\overline{#1}}

\newcommand\varpm{\mathbin{\vcenter{\hbox{%
  \oalign{\hfil$\scriptstyle+$\hfil\cr
          \noalign{\kern-.3ex}
          $\scriptscriptstyle({-})$\cr}%
}}}}
\newcommand\varmp{\mathbin{\vcenter{\hbox{%
  \oalign{$\scriptstyle({+})$\cr
          \noalign{\kern-.3ex}
          \hfil$\scriptscriptstyle-$\hfil\cr}%
}}}}

\definecolor{darkgreen}{HTML}{109930}

\begin{document}

\title{\textbf{\boldmath\Large\mbox{Hidden Sector Custodial Naturalness}\unboldmath}}

\author{Thede de Boer}\email[]{thede.deboer@mpi-hd.mpg.de}
\affiliation{\vspace{0.2cm}Max-Planck-Institut f\"ur Kernphysik, Saupfercheckweg 1, 69117 Heidelberg, Germany}
\author{Manfred Lindner}\email[]{lindner@mpi-hd.mpg.de}
\affiliation{\vspace{0.2cm}Max-Planck-Institut f\"ur Kernphysik, Saupfercheckweg 1, 69117 Heidelberg, Germany}
\author{Andreas Trautner}\email[]{trautner@cftp.ist.utl.pt}
\affiliation{CFTP, Departamento de F\'isica, Instituto Superior T\'ecnico, Universidade de Lisboa, \\ Avenida Rovisco Pais 1, 1049 Lisboa, Portugal}

\begin{abstract}
Custodial Naturalness is a recently introduced idea that combines conformal and scalar-sector custodial symmetry to address the electroweak~(EW) scale hierarchy problem of the Standard Model~(SM). We introduce a new model that realizes Custodial Naturalness without extension of the SM gauge group. The number of new dynamical degrees of freedom is minimized and the custodial symmetry is reduced to $\SO{5}$. This requires a new scalar singlet field that automatically is a good Dark Matter~(DM) candidate, produced via freeze-in with moderate couplings. The most minimal scenario allows the quantum critical generation of the EW scale in a phenomenologically viable way requiring a UV completion at around $10^{11}\,\mathrm{GeV}$. Including ingredients for neutrino mass generation can push this scale to $M_{\mathrm{Pl}}$.
\end{abstract}

\maketitle
\widowpenalty100000\clubpenalty100000
\section{Introduction}
Well-known mechanisms to prevent a sensitivity of the Higgs mass to arbitrary new high scales are supersymmetry or a composite Higgs. However, neither appears the Higgs to be composite, nor have top partners been observed, weighting on naturalness in these proposals. 

Instead of introducing a new scale at which any ``intrinsic'' or ``extrinsic''\footnote{See~\cite{Wells:2025hur,*Wells:2025} for ``intrinsic'' vs.\ ``extrinsic'' hierarchy problems.} hierarchy problem is solved, a different route to a technically natural EW scale could lay in the absence of ``hard'' scales altogether~\cite{Bardeen:1995kv}. Naively, at least one scale in the vicinity of $M_{\mathrm{Pl}}$ should exist, but if gravity itself poses an ``extrinsic'' hierarchy problem is an open question. In particular, the fine-tuning problem may not persist if all scales in nature are generated dynamically.

A well-known mechanism for dynamical scale generation is ``dimensional transmutation'' \'a la Coleman-Weinberg~(CW)~\cite{Coleman:1973jx}, which naturally happens in the quantum critical regime of scalar self-couplings. And indeed, experimental data within $1\sigma$ points to a quantum critical value of the Higgs self-coupling at a high scale~\cite{CMS:2019esx,Hiller:2024zjp,Garces:2025rgn}. Nontheless, CW in the SM is excluded experimentally, as it requires $m_t\lesssim m_Z$ and $m_h\lesssim 10\,\mathrm{GeV}$~\cite{Weinberg:1976pe,Gildener:1976ih}. As a remedy, dimensional transmutation could happen in an extended scalar sector to dynamically induce the EW scale via the Higgs portal~\cite{Hempfling:1996ht,Meissner:2006zh,Espinosa:2007qk,Chang:2007ki,Foot:2007as,Iso:2009ss,Holthausen:2009uc, Farzinnia:2013pga, Hill:2014mqa, Altmannshofer:2014vra}. However, this typically introduces a little hierarchy problem. The mechanism of ``Custodial Naturalness''~\cite{deBoer:2024jne,deBoer:2025oyx} improves on this situation by explaining quantum critical dynamical scale generation including: a technically natural suppression of EW scale, exclusively elementary fields in a perturbative regime, and a marginal top Yukawa coupling as in the SM, without requirement of top partners.

Using the vanilla CW scenario with \U{1} gauge group 
leads to a model with \SO{6} custodial symmetry and, in the minimal case, the same number of free parameters as the SM~\cite{deBoer:2024jne}. Here, we introduce an alternative realization of Custodial Naturalness with minimized field content. This reduces the custodial symmetry to  \SO{5}, while the number of parameters is (at least) one more than in the SM. Without a new gauge group and associated $Z'$ boson, the model becomes much harder to test at colliders. On the flipside, however, there is a new scalar that automatically is a good DM candidate. Refs.~\cite{Ishiwata:2011aa,Kannike:2022pva} have previously studied the same particle content in scale invariant settings
and the intriguing connection to multi-phase criticality~\cite{Kannike:2021iyh,Huitu:2022fcw,Kannike:2022pva}, however, without the extra ingredient of custodial symmetry.

\renewcommand{\arraystretch}{1.15}
\begin{table}[t]
	\begin{tabular}{cccc}
		 Name & \#Gens. & $\SU{3}_\mathrm{c}\!\times\!\SU{2}_\mathrm{L}\!\times\!\U{1}_\mathrm{Y}$ & $\mathbbm{Z}_2$ \\ 
	\hline
    $\phi$ & $1$ & $\left(\rep{1},\rep{1},0\right)$ & $+1$ \\
    $S$ & $1$ & $\left(\rep{1},\rep{1},0\right)$ & $-1$ \\
    \hline
    $N$ & $3$ & $\left(\rep{1},\rep{1},0\right)$ & $+1$
        \\[1pt]
		\hline
	\end{tabular}
	\caption{\label{tab:model}
    Beyond the SM (BSM) fields of ``Hidden Sector Custodial Naturalness.'' The minimal scenario has two real scalars $\phi$ and $S$. The next-to-minimal model additionally contains right-handed Fermions $N$.
    }
\end{table}
\section{Hidden Sector Custodial Naturalness}
The SM is extended by two real scalars $\phi$ and $S$, see Tab.~\ref{tab:model}. The \SO{4} custodial symmetry\footnote{We refer to ``custodial symmetry'' as the full classical symmetry of the scalar potential, cf.~\cite{Sikivie:1980hm}.}
of the SM is promoted to \SO{5}, where $\phi$ and the four real scalars of the SM Higgs doublet $H$ form a $\rep{5}$-plet under custodial symmetry. Spontaneous scale generation \'a la CW requires additional bosonic contributions to the full effective potential, which is the \textit{raison d'\^{e}tre} for the custodial singlet~$S$.

At the scale $\mu=\Lambda_{\mathrm{high}}$, where \SO{5} custodial symmetry and classical scale invariance are realized, the potential  is given by
\begin{equation}\label{eq:CSpotential}\notag
V=\lambda\left(|H|^2+\frac{\phi^2}{2}\right)^2+\frac{\lambda_{\rep{5}S}}{2}\left(|H|^2+\frac{\phi^2}{2}\right)S^2+\frac{\lambda_S}{4!}S^4,
\end{equation}
with real couplings $\lambda$, $\lambda_{\rep{5}S}$, and $\lambda_S$.
As a consequence of classical scale invariance and custodial symmetry, $S$ obeys an exact $\mathbbm{Z}_{2}$ symmetry which stabilizes it as a DM candidate.

\begin{figure}
    \centering
    \includegraphics[width=\linewidth]{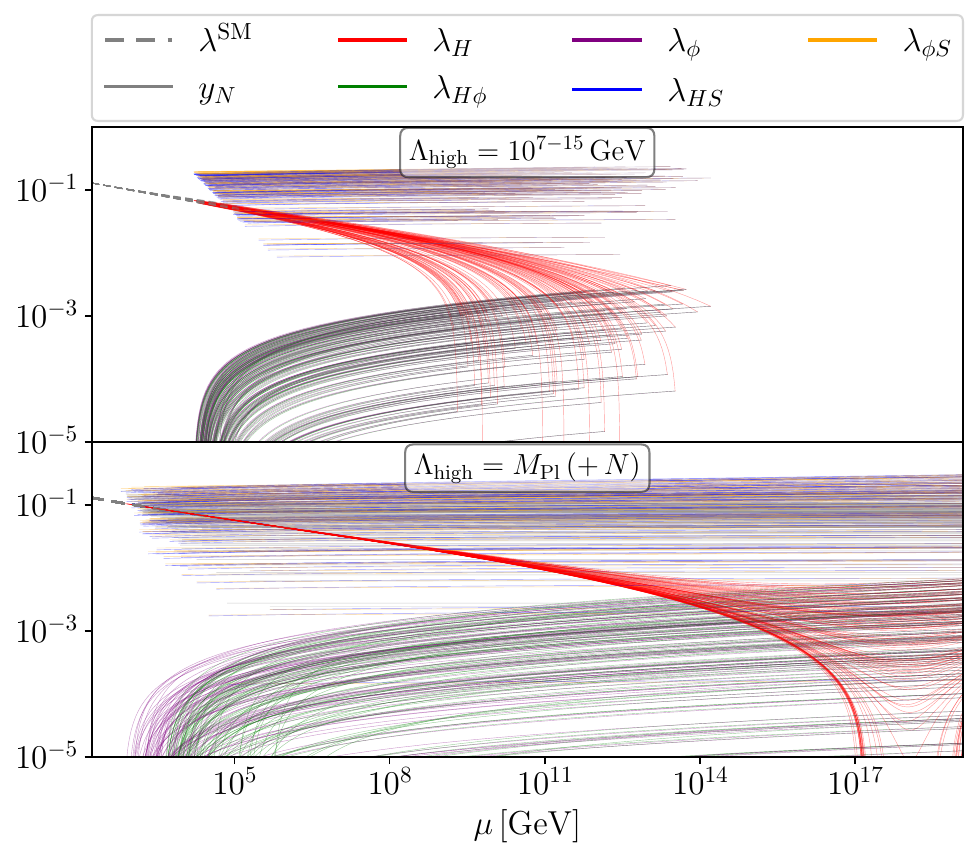}
    \caption{\label{fig:running}%
    Running of scalar couplings below a scale $\Lambda_{\mathrm{high}}$ where 
    $\SO{5}$ custodial symmetry is restored. The top(bottom) plots show the running for the most minimal scenario(extension by three right-handed neutrinos)
    for various parameter points that reproduce the correct EW scale, Higgs and top quark mass.}
\end{figure}
Both, the conformal as well as the scalar sector $\SO{5}$ custodial symmetry are broken by quantum effects and this determines the phenomenology at scales below $\Lambda_{\mathrm{high}}$. Custodial symmetry is explicitly broken by SM gauge and Yukawa interactions -- most dominantly by the top Yukawa coupling $y_t$. The situation is similar to the one discussed in~\cite{deBoer:2024jne,deBoer:2025oyx}. A vacuum expectation value~(VEV), predominantly along the $\phi$-direction, dynamically breaks classical scale invariance and $\SO{5}\to\SO{4}$. The radial mode dilaton $h_\phi$ is the pseudo-Nambu-Goldstone Boson~(pNGB) associated with spontaneous scale symmetry breaking.
The spectrum of pNGB's of custodial breaking consists solely of the very SM-like Higgs boson $h$, while $3$ would-be Nambu-Goldstone modes get eaten by the gauge fields of $\SU{2}_{\mathrm{L}}\times \U{1}_{\mathrm{Y}}$ in the cause of EWSB.
$S$ is rather independent of the other scalars as it is not a pseudo-Goldstone mode and does not obtain a VEV, $\langle S\rangle=0$, hence, also does not mix with the other scalars.

A simple extension of the most minimal model is to add 3 generations of right-handed (RH) neutrinos $N$, see Tab.~\ref{tab:model}, which introduces a BSM source of explicit custodial symmetry breaking given by the Yukawa couplings 
\begin{equation}\label{eq:YN}
\mathcal{L}_\text{Yuk}=\frac{1}{2}y_N\phi NN+\mathrm{h.c.}
\end{equation}
This results in Lepton number violating $N$ masses of the size of the intermediate scale $\langle\phi\rangle$ and light neutrino mass generation via a type I seesaw~\cite{Minkowski:1977sc, *Mohapatra:1979ia, *Yanagida:1980xy, *Gell-Mann:1979vob, *Glashow:1979nm}. In this extended scenario, the $\mathbbm{Z}_2$ symmetry of $S$ must be imposed to ensure DM stability.

Below $\Lambda_{\mathrm{high}}$ the custodial symmetry does not hold and we parametrize the potential as
\begin{equation}\label{eq:potential}
\begin{split}
    V~=~&\lambda_H|H|^4+\lambda_{H\phi}|H|^2\phi^2+\frac{\lambda_\phi}{4}\phi^4\\
    &+\frac{\lambda_{HS}}{2}|H|^2S^2+\frac{\lambda_{\phi S}}{4}\phi^2S^2+\frac{\lambda_S}{4!}S^4\;.
\end{split}
\end{equation}
The dominant custodial breaking by $y_t$ drives $\lambda_H$ to a large value while
$\lambda_\phi\sim\lambda_{H\phi}$ and $\lambda_{HS}\sim\lambda_{\phi S}$ stay close to each other. The running for several phenomenologically viable initial values is shown in Fig.~\ref{fig:running}. As required by the CW mechanism, $\lambda_\phi$ runs to quantum critical small values at low scales. This results in a flat direction of the potential pointing predominantly along the $\phi$-field direction, ensuring a hierarchy of VEVs
\begin{equation}\label{eq:vevs}
v_\phi\equiv\langle \phi\rangle\quad\gg\quad
v_H\equiv\sqrt{2}\langle H\rangle\;.
\end{equation}
This establishes $v_\phi$ as intermediate scale of spontaneous scale and custodial symmetry violation, while the EW scale is custodially suppressed.

The scalar masses are approximately given by\footnote{These are valid around the CW scale. More reliable, renormalization scale independent expressions 
are stated in App.~\ref{sec:Appendix}.}
\begin{equation}\label{eq:masses}
m_{h_\phi}\approx\beta_{\lambda_\phi}\,v_\phi^2\approx
\frac{\lambda_{\phi S}^2-8 y_N^4}{32\pi^2}\,v_{\phi}^2,
\end{equation}\vspace{-0.57cm}
\begin{equation}
m_h^2\approx2\left(\lambda_\phi-\lambda_{H\phi}\right)v_\phi^2,
\quad
m_S^2\approx\frac12 \lambda_{\phi S}v_\phi^2,
\end{equation}
where $\beta_{g_i}$ denotes the beta function of a coupling $g_i$. 
Custodial symmetry ensures $|\lambda_\phi-\lambda_{H\phi}|\ll1$, thereby implying custodial suppression of the EW scale $v_H$, which obeys the SM relation $v_H^2\approx m^2_h/2\lambda_H$.

A sufficient amount of custodial symmetry breaking is crucial here to make $\lambda_\phi-\lambda_{H\phi}$
\textit{large enough} and of the right sign. Otherwise, the EW scale is too suppressed or not generated at all. The leading SM contribution to this splitting is given by 
\begin{equation}\label{eq:betaSM}
	\beta_{\lambda_{H\phi}}-\beta_{\lambda_\phi}\biggr|_\mathrm{SM}\simeq\frac{\lambda_{H\phi}}{16\pi^2}\left[-\frac92 g_L^2-\frac32 g_Y^2+12 \lambda_H +6 y_t^2\right].
\end{equation}
Including $\phi$-$N$ Yukawa couplings \eqref{eq:YN} adds a contribution 
\begin{equation}\label{eq:betaBSM}
	\beta_{\lambda_{H\phi}}-\beta_{\lambda_\phi}\biggr|_{y_N}\simeq\frac{y_N^4}{4\pi^2}\;.
\end{equation}
The SM contribution by itself only yields a realistic EW scale \textit{if} the custodially symmetric boundary conditions are required at a specific scale. 
That is, in the minimal model without $y_N$ the possible values of $\Lambda_{\mathrm{high}}$
are predictions of a correct low energy phenomenology.\footnote{The same is true for the \SO{6} Custodial Naturalness model~\cite{deBoer:2024jne}, where the scale predicted in this way is $\Lambda_{\mathrm{high}}\sim10^{11}\,\mathrm{GeV}$~\cite{deBoer:2025oyx}.}
As visible form Fig.~\ref{fig:running} (top) the minimal scenario is only viable for $\Lambda_{\mathrm{high}}\in[10^{9},10^{13}]\,\mathrm{GeV}$. 
If one allows for an additional source of custodial symmetry breaking, e.g.\ including $y_N$, $\Lambda_{\mathrm{high}}$ can be pushed higher, e.g.\ $\Lambda_{\mathrm{high}}\sim M_{\mathrm{Pl}}$ as imposed in Fig.~\ref{fig:running} (bottom).
The top quark mass is sensitive to the precise value of $\Lambda_{\mathrm{high}}$, implying that a more precise measurement of $M_t$ can further constrain this scale.

\begin{figure}
    \centering
    \includegraphics[width=0.5\textwidth]{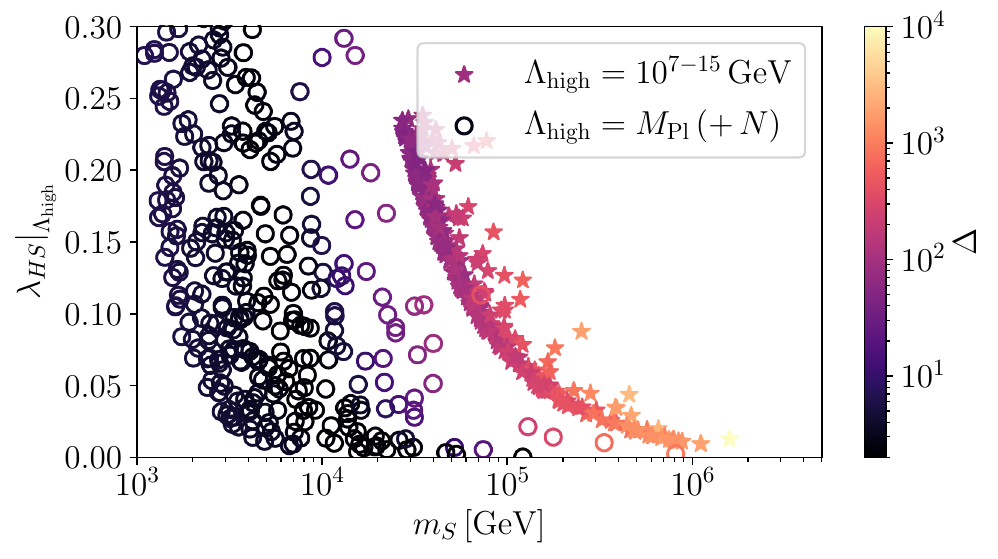}
    \caption{\label{fig:lamms}
    Viable parameter space of the model and required fine tuning in order to reproduce the correct EW scale, Higgs and top mass. Stars denote the minimal scenario, while circles correspond to the case which includes RH neutrinos.}
\end{figure}
\section{Numerical analysis}
To explore the parameter space we perform a numerical scan following the procedures described in~\cite{deBoer:2024jne,deBoer:2025oyx}. We use a numerical evaluation of the 1-loop effective potential and two-loop running in $\bar{\mathrm{MS}}$ scheme with beta functions computed with~\texttt{PyR@TE}~\cite{Sartore:2020gou}.
The intermediate scale $\phi_0$ and matching scale to the SM $\mu_0$ are iteratively determined as self-consistent solution to 
Eq.~\eqref{eq:phi0_long}, see Appendix for definitions.
At $\mu_0$ we numerically minimize the 1-loop effective potential to compute the VEVs $v_\phi$, $v_H$, scalar masses and couplings. These are predictions of each model point. Integrating out all heavy states, we match to the SM 1-loop effective potential below which we 2-loop evolve to the top mass (dilaton contributions are negligible). We select phenomenologically viable points that reproduce the electroweak scale,  physical Higgs and top masses. We use~\cite{Buttazzo:2013uya} to relate the top pole mass $M_t$ and $\bar{\mathrm{MS}}$ top Yukawa coupling.

For both model variations the starting value (before running up the first time) for $\tilde{\mu}_0$ is sampled in $\tilde\mu_0\in[500,10^6]\,\mathrm{GeV}$, initial portal coupling $\lambda_{\phi S}\bigr|_{\tilde\mu_0}\in[0,0.2]$, and we fix $\lambda_{S}\bigr|_{\tilde\mu_0}=0.5$. In the minimal model we randomly sample the scale of custodially symmetric boundary conditions in $\Lambda_\mathrm{high}\in[10^6,10^{17}]\,\mathrm{GeV}$. By contrast, including RH neutrinos we fix $\Lambda_\mathrm{high}=M_{\mathrm{Pl}}$ while sampling the additional source of custodial symmetry violation within $y_N\bigr|_{\tilde\mu_0}\in[0,0.4]\cdot\sqrt{\lambda_{\phi S}}\bigr|_{\tilde\mu_0}$.

This results in a set of phenomenologically viable parameters that reproduce the EW VEV, Higgs and top mass. The fine tuning for all parameter points is computed with a Barbieri-Giudice-type~\cite{Barbieri:1987fn} measure $\Delta=\max_{g_i}\;\Delta_{g_i}^{\langle H\rangle}-\Delta_{g_i}^{\langle\phi\rangle}$ defined in~\cite{deBoer:2024jne,deBoer:2025oyx}. 
The phenomenologically viable parameter space and required fine tuning for both model variations is shown in Fig.~\ref{fig:lamms}. Fine-tuning is absent in the case of a high-scale $\Lambda_\mathrm{high}$, but moderately present in the case without RH neutrinos. The reason is that due to the higher intermediate scale, the Higgs mass is sensitive to the cancellation among the SM parameters in (\ref{eq:betaSM}).

\begin{figure}
    \includegraphics[width=0.5\textwidth]{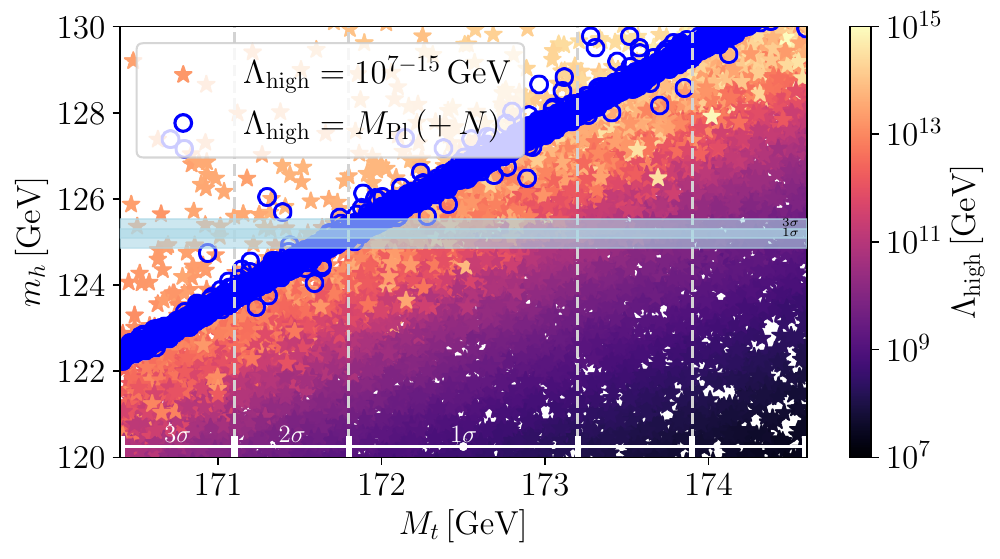}%
    \caption{\label{fig:Mtmh}%
    Correlation of the top pole mass $M_t$, Higgs mass $m_h$ and the scale of custodial symmetry $\Lambda_{\mathrm{high}}$ for the minimal case (stars) as well as for the case with three RH neutrinos (blue circles). All displayed points reproduce the correct EW scale.
    } 
\end{figure}

There is a correlation between Higgs and top mass shown in Fig.~\ref{fig:Mtmh}.
For the scenario with Planck-scale $\Lambda_\mathrm{high}$ (blue circles) this gives the same prediction as already found in the $\SO{6}$ case~\cite{deBoer:2024jne,deBoer:2025oyx}: The top pole mass $M_t$ should reside at the lower end of its currently allowed $1\sigma$ interval. For the case without a new source of custodial breaking, $\Lambda_\mathrm{high}$ is allowed to vary over a certain range of scales. The prediction for the top mass correlates with $\Lambda_\mathrm{high}$ as shown by the color coding in Fig.~\ref{fig:Mtmh}. Measuring $M_t$ more precisely would allow to refine the  value of $\Lambda_\mathrm{high}$, where higher top masses correspond to lower UV completion scales.

\begin{figure}
    \includegraphics[width=0.5\textwidth]{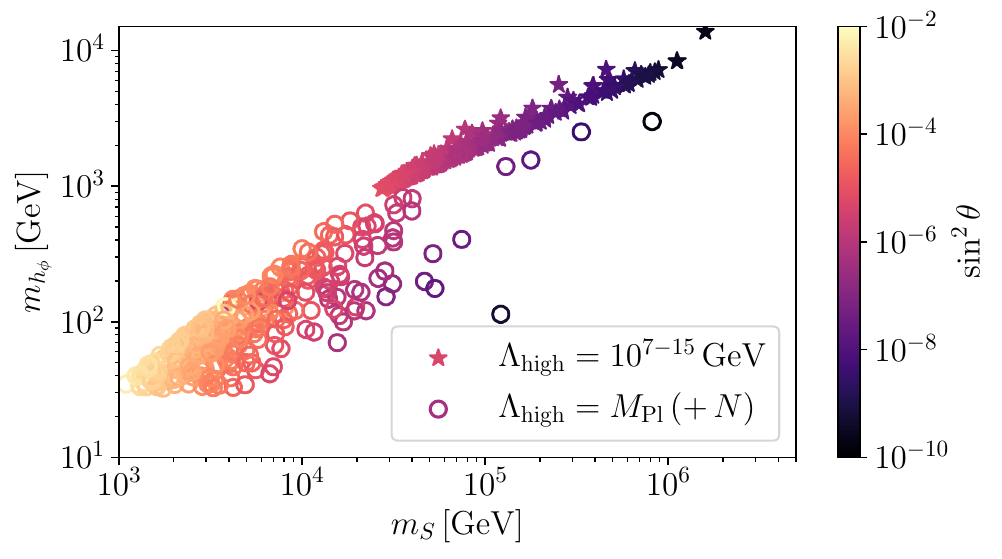}
    \caption{\label{fig:mSmhphi}
    Correlation of the dilaton $m_{h_{\phi}}$ and DM candidate $m_S$ masses, also showing the corresponding Higgs-dilaton mixing angle $\sin\theta$.
    }
\end{figure}

\section{Phenomenology}\enlargethispage{1.0cm}
\subsection{Masses and mixing}
We use a fully numerical minimization of the 1-loop effective potential and numerical evaluation of all masses and mixing angles, thereby also confirming the analytic approximations of Eq.~\eqref{eq:masses} and App.~\ref{sec:Appendix}.
The resulting masses and Higgs-dilaton mixing angle $\theta$ for all viable parameter points are shown in Fig.~\ref{fig:mSmhphi}. In the minimal scenario, scalar masses are heavier and range from $20\,\mathrm{TeV}\lesssim m_S\lesssim2\,\mathrm{PeV} $ for the DM candidate and $1\,\mathrm{TeV}\lesssim m_{h_\phi}\lesssim10\,\mathrm{TeV}$ for the dilaton. In the scenario with RH neutrinos and a high UV completion scale, viable points have $1\,\mathrm{TeV}\lesssim m_S\lesssim10^2\,\mathrm{TeV}$ and $30\,\mathrm{GeV}\lesssim m_{h_\phi}\lesssim1\,\mathrm{TeV}$. 

Including RH neutrinos, the neutrino mass matrix after symmetry breaking is given by
\begin{equation}
    M_\nu=\left(\begin{matrix}
        0&\frac{v_H}{2\sqrt2}y_D\\
        \frac{v_H}{2\sqrt2}y_D& y_N v_\phi
    \end{matrix}
    \right)\;,\label{eq:Neutrino mass matrix}
\end{equation}
where $y_D$ are the usual Dirac-type Yukawa couplings of the Lepton doublet to RH neutrinos. Eq.~(\ref{eq:Neutrino mass matrix}) is the same as in the type I seesaw mechanism. Assuming $y_Nv_\phi\sim\mathrm{TeV}$, realistic active neutrino Majorana masses are be obtained with $y_D\sim10^{-5}$.

\subsection{Collider searches}\enlargethispage{1.0cm}
The models discussed here are much harder to test than in the case of \SO{6} Custodial Naturalness~\cite{deBoer:2024jne, deBoer:2025oyx} due to the absence of a $Z'$ gauge boson.  Furthermore, there is no BSM source of custodial symmetry violation in the minimal case, implying that there are no observable effects on electroweak precision. Including $y_N$ does not change this much, as the only way in which the EW sector is
affected is through Higgs-dilaton mixing.
The leading signatures in the present case are very similar to the SM+scalar singlet extension, see e.g.~\cite{Robens:2015gla, *Godunov:2015nea, *Falkowski:2015iwa, *Arcadi:2019lka}. In our case the dilaton $h_\phi$ (not $S$) 
corresponds to the singlet scalar target of simplified models. 
The Higgs-dilaton mixing angle is approximately 
\begin{equation}\label{eq:mixingangle}
\tan\theta\approx
\frac{2\left(\lambda_{H\phi}-\lambda_\phi-\frac{\lambda_{HS}\lambda_{\phi S}}
{64\pi^2}\right)}{m_h^2-m_{h_\phi}^2}v_\phi v_H\;.
\end{equation}
The latest constraints on this can be found in \cite{Robens:2024wbw, Abidi:2025dfw} and references therein. They are never more restrictive than $\sin\theta\lesssim0.1$, which is true for all of our parameter points, see Fig.~\ref{fig:mSmhphi}, even in the ``resonant'' region where $m_h^2\sim m_{h_\phi}^2$ and Eq.~\eqref{eq:mixingangle} does not hold.
Constraints on the mixing angle and other scalar couplings will be considerably strengthened at HL-LHC, as well as at future colliders like FCC-ee/hh and Higgs factories like the ILC~\cite{Abidi:2025dfw}. 

In contrast to the singlet scalar extension, the dilaton $h_\phi$ has additional couplings to SM gauge bosons originating from the conformal trace anomaly~\cite{Goldberger:2007zk,Chacko:2012sy,Bellazzini:2012vz}. Effects of these couplings are suppressed $\propto v_{H}/v_{\phi}$ and, therefore, currently numerically irrelevant. Ultimately, however, these could be instrumental to distinguish a dilaton from other scalar extensions~\cite{Ahmed:2015uqt,Ahmed:2019csf}.

The DM candidate $S$ couples to the SM only through its coupling with the Higgs $\lambda_{HS}$ (and subdominant Higgs-dilaton mixing effects), and can only be produced in pairs owing to the preserved $\mathbbm{Z}_2$ symmetry. Together with masses of $S$ that reach up to the $\mathrm{PeV}$ scale this realizes a ``nightmare'' scenario for collider detectability of~DM.

\interfootnotelinepenalty=10000
\begin{figure*}
    \centering
    \includegraphics[width=0.5\textwidth]{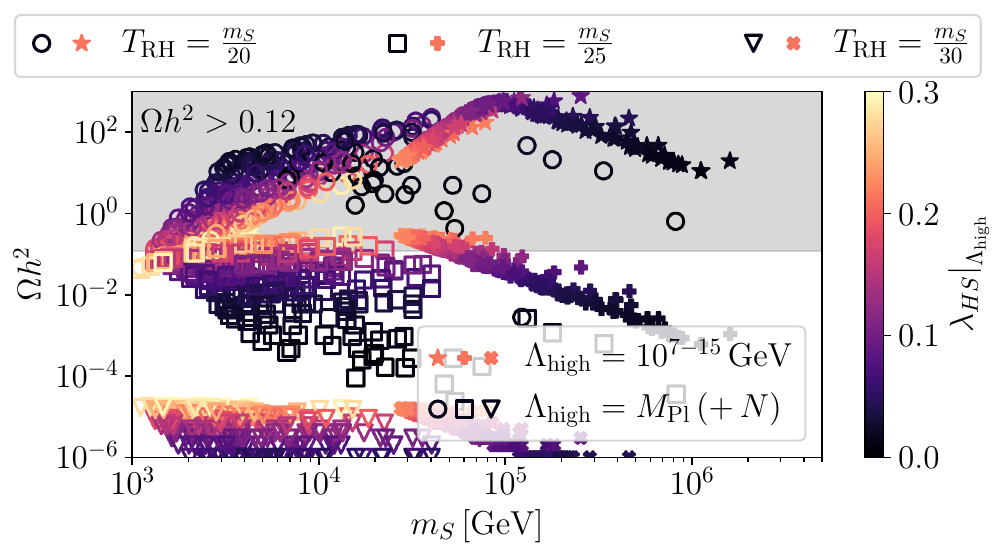}%
    \includegraphics[width=0.5\textwidth]{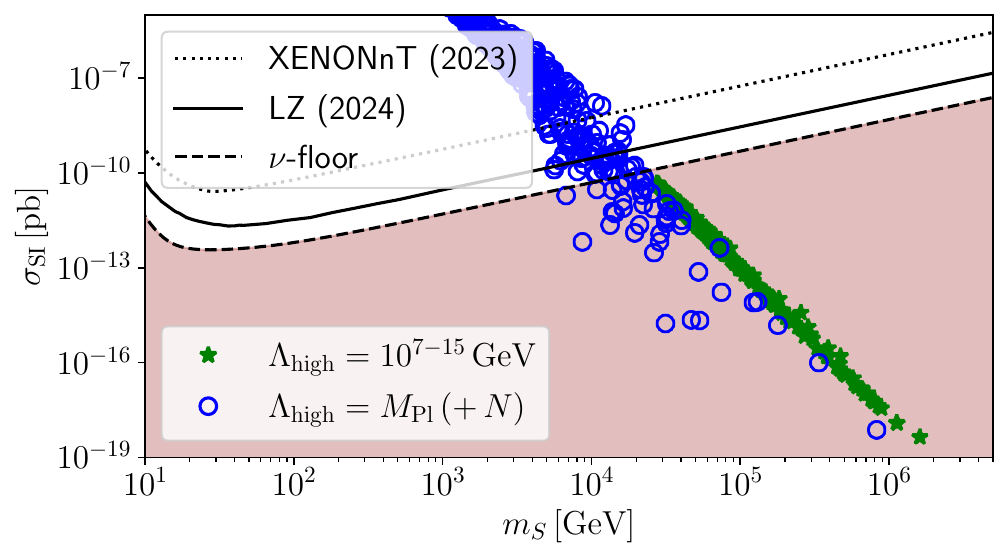}
    \caption{\label{fig:relic_density}
    Left: DM relic density as a function of DM mass $m_S$ and reheating temperature $T_{\mathrm{RH}}$ for both models. Right: Corresponding spin-independent direct detection cross section for all otherwise viable points that also saturate the observed relic density (for each point we assume that $T_{\mathrm{RH}}$ takes a value such that $\Omega h^2$ is within the observed bounds). We also show limits from XENON1T~\cite{XENON:2023cxc} and LZ~\cite{LZ:2024zvo} (linearly extrapolated to large $m_S$).
    }
\end{figure*}
\subsection{Dark Matter}\enlargethispage{1cm}
In the early Universe, our scenarios feature a period of strong supercooling and thermal inflation followed by a first-oder phase transition, see e.g.\ the recent~\cite{Liu:2024fly}. The subsequent reheating reaches temperatures $T_\mathrm{RH} \lesssim m_S$~\cite{Liu:2024fly}.\footnote{Dimensional analysis gives $T_{\mathrm{RH}}\lesssim(\Delta V)^{1/4}\approx m_S$.
The exact value of $T_{\mathrm{RH}}$ depends on details of the reheating process, such as whether it is slow or instantaneous.
\label{fot:TRH}
} 
Hence, regular freeze-out production of the DM candidate $S$ is hampered. Nonetheless, a sufficient DM relic density can be produced via freeze-in at moderate couplings~\cite{Cosme:2023xpa,Cosme:2024ndc,Arcadi:2024obp,Arcadi:2024wwg,Lebedev:2024mbj}. In this case, stronger couplings to the SM bath (as compared to an ordinary FIMP) allows for production of DM from the tail of the Maxwell-Boltzmann distribution, if $m_S$ is not too far above $T_{\mathrm{RH}}$. The yield of this production mechanism, hence, is exponentially sensitive to $T_{\mathrm{RH}}/m_S$. Owing to the complicated dynamics of the phase transition we can currently not compute $T_{\mathrm{RH}}$, hence, take it as a free parameter (see footnote \ref{fot:TRH}). 

We have calculated the DM relic density using \texttt{micrOMEGAs}~$6.0.5$~\cite{Alguero:2023zol} with model files generated using \texttt{SARAH-4.15.2}~\cite{Staub:2013tta}. The results for all viable parameter points are shown in Fig.~\ref{fig:relic_density}~(left). For each value of $m_S$ there exists a reheating temperature in the narrow range $20\lesssim m_S/T_{\mathrm{RH}}\lesssim30$ for which the correct DM relic density is obtained. In Fig.~\ref{fig:relic_density}~(right) we show the corresponding spin-independent direct detection cross section and current constraints from XENON1T~\cite{XENON:2023cxc} and LZ~\cite{LZ:2024zvo}. Hence, $S$ is a viable DM candidate that can further be constrained by direct detection searches. Explaining the coincidence between DM and baryon relic density in this class of models requires a computation of $T_{\mathrm{RH}}$ from first principles. This is currently out of reach, just like computing the exact gravitational wave signals from the expected bubble collisions after the strong first-order phase transition, see e.g.~\cite{Kierkla:2025vwp}.
The diminished collider-detectability motivates further work in this direction because it renders gravitational wave searches and DM direct detection the most promising ways to test Hidden Sector Custodial Naturalness.

\section{Conclusions}
We have presented a new minimal (in terms of field content) realization of the idea of Custodial Naturalness, where the Higgs boson arises as a pNGB of an explicitly and spontaneously broken extended scalar sector custodial symmetry. This explains the suppression of the EW scale in a technically natural way.
In the present case, the extended high-scale custodial symmetry is given by \SO{5}. This requires two BSM scalar fields, both singlets under the SM gauge group. On the one hand, there is the dilaton as pNGB of spontaneous and explicit scale symmetry breaking, which can be searched for at future colliders and specifically Higgs-factories through Higgs-dilaton mixing. On the other hand, there is the DM candidate $S$ which is produced through Boltzmann-suppressed freeze-in at moderate couplings, and whose properties can further be constrained by direct detection experiments.  
Other signatures of the model include gravitational waves produced after a strongly supercooled first-order phase transition, details of which have to be worked out. 

There are further theoretical questions about the mechanism of Custodial Naturalness. Namely, how scalar (self-)couplings can be made asymptotically free or safe and, more generally, how custodially symmetric boundary conditions can be obtained in UV embeddings such as grand unified theories or theories where the Planck mass itself is spontaneously generated by a breaking of conformal invariance.
This connects to the important question of whether or not the dynamical generation of all scales can prevent the existence of an extrinsic hierarchy problem to the Standard Model.

\acknowledgments
AT is supported by the Portuguese Funda\c{c}\~ao para a Ci\^encia e a Tecnologia (FCT) in the project \href{https://doi.org/10.54499/2023.06787.CEECIND/CP2830/CT0005}{2023.06787.CEECIND} and in parts through contract \href{https://doi.org/10.54499/2024.01362.CERN}{2024.01362.CERN}, partially funded through POCTI (FEDER), COMPETE, QREN, PRR, and the EU.

\bibliographystyle{utphys}
\bibliography{bib.bib}

\onecolumngrid
\vspace{1cm}
\hrule
\appendix
\section{Supplemental Material \label{sec:Appendix}}
\noindent\enlargethispage{1cm}
The tree-level potential has already been stated in Eq.~\eqref{eq:potential}. The one-loop effective potential is computed as 
\begin{equation}
V_\text{eff}=V_\text{tree}+ \sum_i \frac{n_i (-1)^{2 s_i}}{64\pi^2}m_{i,\text{eff}}^4\left[\ln\left(\frac{m_{i,\text{eff}}^2}{\mu^2}\right)-C_i\right]\;,
\end{equation}
where $n_i$ denotes the number of real degrees of freedom in each field $i$, $(-1)^{2s_i}=\varpm1$ for bosons(fermions), and $C_i=\frac{5}{6}\left(\frac{3}{2}\right)$ for vector bosons(scalars or fermions). The effective tree level squared masses $m_{i,\mathrm{eff}}^2$ can be obtained as the eigenvalues of 
\begin{equation}
\begin{pmatrix}
2 \lambda_{H\phi} H_b^2 + 3 \lambda_\phi \phi_b^2 & 2 \sqrt{2} \lambda_{H\phi} H_b \phi_b \\
2 \sqrt{2} \lambda_{H\phi} H_b \phi_b & 6\lambda_H H_b^2 + \lambda_{H\phi} \phi_b^2
\end{pmatrix}\;,
\end{equation}
as well as by the eigenvalues of 
\begin{equation}
\mathrm{diag}\{2 \lambda_H H_b^2 + \lambda_{H\phi} \phi_b^2,\;2 \lambda_H H_b^2 + \lambda_{H\phi} \phi_b^2,\;2 \lambda_H H_b^2 + \lambda_{H\phi} \phi_b^2,\;\lambda_{HS} H_b^2 +\frac12 \lambda_{\phi S} \phi_b^2\}\;,
\end{equation}
which we have obtained from the tree-level potential using the background fields $H=H_b$ and $\phi=\phi_b$. In case we include RH neutrinos $N$, their effective masses are given by $m_{\psi,\text{eff}}=y_N\phi_b$. For simplicity, we assume that a single entry in the Yukawa matrix $y_N$ dominates and we denote this value as $y_N$ while neglecting the other entries. The effective potential above forms the basis of all of our numerical evaluations. 

For an analytic insight, we can follow the procedure already used in~\cite{deBoer:2024jne,deBoer:2025oyx} and define an effective potential 
\begin{equation}\label{eq:potEFF}
V_\mathrm{EFT}(H_b):=V_\mathrm{eff}\left(H_b,\tilde{\phi}(H_b)\right)\,,
\end{equation}
where $\tilde{\phi}(H_b)$ is implicitly defined via 
\begin{equation} 
	\left.\frac{\partial V_\text{eff}}{\partial \phi_b}\right|_{\phi_b=\tilde{\phi}(H_b)}=0\,.
\end{equation}
The VEV of $\phi$ in the limit $\langle\phi\rangle\gg\langle H\rangle$ is approximated by $\phi_0:=\tilde\phi(H_b=0)$ which is given by
\begin{equation}\label{eq:phi0_long}
\begin{split}
    \ln\left(\frac{\phi_0^2}{\mu^2}\right)=&-\frac{32 \pi ^2 \lambda_\phi+\frac{1}{2} \lambda_{\phi S}^2 \left[\ln \left(\frac{\lambda_{\phi S}}{2}\right)-1\right]+8 \lambda_{H\phi}^2 \left[\log \left(\lambda_{H\phi}\right)-1\right]-4 y_N^4 \left[\ln \left(y_N^2\right)-1\right]}{\frac{\lambda_{\phi S}^2}{2}+8\lambda_{H\phi}^2-4 y_N^4}\\
    \approx&-\frac{64\pi ^2 \lambda_\phi}{\lambda_{\phi S}^2}-\ln \left(\frac{\lambda_{\phi S}}{2}\right)+1\;.
\end{split}
\end{equation}
Expanding $V_\mathrm{EFT}(H_b)$ in $H_b/\phi_0\ll1$ at the quadratic order in $H_b$ one finds 
\begin{equation}
\begin{split}
V_\text{EFT}~\approx&~
\left\{\lambda_{H\phi}-\frac{\lambda_{HS}\lambda_{\phi S}}{16\lambda_{H\phi}^2+\lambda_{\phi S}^2-8 y_N^4}\left[\lambda_\phi-\frac{2\,y_N^4}{16\pi^2} \ln\left(\frac{2\,y_N^2}{\lambda_{\phi S}}\right)\right]\right\}\phi_0^2 H_b^2
~\approx~\left(\lambda_{H\phi}-\lambda_\phi\right)\phi_0^2 H_b^2\;.
\end{split}
\end{equation}
This expression is RG-invariant and exposes how custodial symmetry breaking terms generate the Higgs quadratic term by the differential running of \mbox{$\lambda_{H\phi}-\lambda_\phi$}.

To analyze the matching to the SM effective potential a different expansion is useful. For this, we introduce an artificial expansion parameter $\epsilon$ such that 
\begin{equation}
	\frac{H_b}{\phi_0}\to\epsilon\frac{H_b}{\phi_0},\qquad \lambda_{H\phi}\to\epsilon^2\lambda_{H\phi}.
\end{equation}
This ensures that expanding around $\epsilon\to0$ we obtain
\begin{equation}
	\frac{\phi_0}{H_b}\to \infty,\qquad\frac{\lambda_{H\phi}}{\lambda_H}\to0,\qquad\text{while keeping}\qquad\lambda_{H\phi} \phi_0^2=\lambda_H H_b^2\quad\text{fixed}.
\end{equation}
This is a 't Hooft-Veneziano-like expansion which automatically realizes the leading-order Gildener-Weinberg condition $\lambda_{H\phi} \phi^2=\lambda_H H^2$. This condition is obtained by assuming $\lambda_{HS}\approx\lambda_{\phi S}>0$ and $\lambda_{H\phi}<0$ such that at a scale $\mu=\mu_{GW}$ the flat direction is given by
\begin{equation}
    \phi=\sqrt{\frac{2\lambda_H}{2\lambda_H-\lambda_{H\phi}}}r,\quad H=\sqrt{\frac{-\lambda_{H\phi}}{2\lambda_H-\lambda_{H\phi}}}r,\quad\lambda_\phi=\frac{\lambda_{H\phi}^2}{\lambda_H}\;,
\end{equation}
where $r=\sqrt{H^2+\phi^2}$. The matching to the SM now most conveniently is done at a scale $\mu=\mu_0:=\sqrt{\lambda_{\phi S}/(2\mathrm{e})}\,\phi_0$ in order to avoid large logarithms. Expanding the effective potential Eq.~\eqref{eq:potEFF} in powers of $\epsilon$ up to $\epsilon^4$ 
at the scale $\mu=\mu_0$ we find
\begin{equation}
    V_\text{EFT}=-\frac{\frac12 \lambda_{\phi S}^2-4y_N^4}{256\pi^2}\phi_0^4+\lambda_{H\phi}\phi_0^2H_b^2+\lambda_HH_b^4
+\sum_i\frac{n_i
(-1)^{2s_i}}{64\pi^2}m_{i,\text{eff}}^4\left[\ln\left(\frac{m_{i,\text{eff}}^2}{\mu_0^2}\right)-C_i\right]
-\frac{y_N^4\lambda_{HS}^2}{8\pi^2\left(\lambda_{\phi S}^2-8
y_N^4\right)}H_b^4.
\end{equation}
Here, the sum over $i$ is understood as a sum over the SM effective masses assuming a tree level potential given by $V=\lambda_{H\phi}\phi_0^2|H|^2+\lambda_H|H|^4$.
This expression is suitable to match to the SM one-loop effective potential. The masses we obtain from this expression are given by
\begin{align}
m_S^2~&=~\frac12\lambda_{HS} v_H^2 +\frac12 \lambda_{\phi S} v_\phi^2\approx\frac12 \lambda_{\phi S} v_\phi^2\;,& \\
m_{h_\phi}~&=~\frac{\frac{\lambda_{\phi S}^2}{2}+8\lambda_{H\phi}^2-4 y_N^4}{16\pi^2}v_\phi^2\approx\frac{\frac{\lambda_{\phi S}^2}{2}-4 y_N^4}{16\pi^2}v_\phi^2\approx\beta_{\lambda_\phi}v_\phi^2~\;,& \\
m_h^2~&=~-2\left\{\lambda_{H\phi}-\frac{\lambda_{HS} \frac{\lambda_{\phi S}}{2}}{8\lambda_{H\phi}^2+\frac{\lambda_{\phi S}^2}{2}-4 y_N^4}\left[\lambda_\phi-\frac{2 y_N^4}{16\pi^2} \ln \left(\frac{2 y_N^2}{\lambda_{\phi S}}\right)\right]\right\}v_\phi^2
\approx2\left(\frac{\lambda_{HS}\lambda_\phi}{\lambda_{\phi S}}-\lambda_{H\phi}\right)v_\phi^2
\approx2\left(\lambda_\phi-\lambda_{H\phi}\right)v_\phi^2
\;.&
\end{align}
For the Higgs-dilaton mixing angle one finds
\begin{equation}
\tan\theta~\approx~\frac{2\left\{\lambda_{H\phi}-\frac{\lambda_{HS} \frac{\lambda_{\phi S}}{2}}{8\lambda_{H\phi}^2+\frac{\lambda_{\phi S}^2}{2}-4 y_N^4}\left[\lambda_\phi-\frac{2 y_N^4}{16\pi^2} \ln \left(\frac{2 y_N^2}{\lambda_{\phi S}}\right)\right]+\frac{\lambda _{HS} \lambda_{\phi S}}{64\pi^2}\right\}}{m_h^2-m_{h_\phi}^2}v_H v_\phi
~\approx~
\frac{2\left(\lambda_{H\phi}-\lambda_\phi-\frac{
\lambda_{HS}\lambda_{\phi S}}
{64\pi^2}\right)}{m_h^2-m_{h_\phi}^2}v_\phi v_H\;.
\end{equation}
\twocolumngrid

\end{document}